\documentstyle[graphicx,aps,preprint]{revtex}
\begin{document}
\vspace{1.0ex}
%{\hskip 12.0cm} hep-ph/98mmnnn\\
\vspace{6ex}
\begin{center}        
{\LARGE \bf Decay $\Lambda_b \rightarrow p l \bar \nu$ in QCD sum rules}\\
\vspace{5ex}
{ Chao-Shang Huang$^{a}$, Cong-Feng Qiao$^{b,c}$, and Hua-Gang Yan$^{a}$}\\
\vspace{3ex}
{\it  $^a$ Institute of Theoretical Physics, Academia Sinica}\\
{\it P.O. Box  2735,  Beijing  100080, China
\footnote{Corresponding address}}\\
{\it $^b$ CCAST (World Laboratory) P.O. Box 8730, Beijing 100080, China}\\     
{\it $^c$ ICTP, P.O. Box 586, 34100 Trieste, Italy}\\
\vspace{8.0ex}
{\large \bf Abstract}\\
\vspace{4ex}
\begin{minipage}{130mm}
The $\Lambda_b \rightarrow p$ semileptonic decay is analyzed by using QCD
sum rules within the framework of heavy quark effective theory.
The Isgur-Wise function of $\Lambda_b \rightarrow p l \bar\nu$ has been
calculated. The decay width is given. 
\par
\vspace{0.5cm}
\end{minipage}
\end{center}

\newpage

The semileptonic decays of heavy hadrons have been widely investigated as a 
testing tool for the Standard Model(SM). They can reveal some important features
of the physics of heavy quarks with some uncertainties, which are rooted in the
nonperturbative aspects of the strong interactions(QCD). Employing
the QCD sum rule method$\cite{svz}$, we can find a way to estimate the nonperturbative
effects, hence can extract some Cabbibo-Kobayashi-Maskawa(CKM) matrix
elements by comparing experiments with theoretical calculations.

For the heavy hadrons containing one single heavy quark, an effective
theory of QCD based on the heavy quark symmetry in the heavy quark
limit $\cite{iml}$, the so-called heavy quark effective theory(HQET),
has been proposed$\cite{gl}$. For heavy quark to heavy quark transitions,
both the classification of the weak decay
form factors of heavy hadrons$\cite{imn,mrrr}$ and the computation of the 
form factors by using sum rules can be greatly simplified in HQET
$\cite{all,neur4,liul}$. For the
 heavy quark to light quark semileptonic
decay mode, the QCD sum rule method has yet not been used within the
framework of HQET. In this letter we shall do such a study.

For heavy to light transitions, one may make use of the heavy quark symmetry for
the heavy quark (b or c). It has been shown that the heavy quark symmetry restricts
 the number of form factors for the heavy baryonic transition $\Lambda_Q\rightarrow
$light spin-1/2 baryon to two$\cite{mrrr}$. That is, in the heavy quark limit, 
the hadronic matrix element of the transition $\Lambda_b \rightarrow
p $ is characterized generally by two form factors $F_1$ and $F_2$,
\begin{equation}
\begin{array}{lll}
<\Lambda_b(v)|\bar{b}\Gamma u|p(P)>&=&
\bar{u}_{\Lambda_b}(v)\Gamma (F_1+F_2 \not\! v ) u_{p}(P),
\end{array}
\end{equation}
where $v$ denotes the four-velocity of $\Lambda_b$, $P$ denotes the
four-momentum of $p$, and $\Gamma$ is an arbitrary Dirac matrix ($\Gamma=\gamma_{\mu}
(1-\gamma_5)$ for $\Lambda_b\rightarrow p l\bar{\nu}$) . Hereafter 
we focus on the determination of $F_1$ 
and $F_2$ by considering the three-point correlators of baryonic currents
\begin{equation}
\begin{array}{lll}
\tilde{j}^v=\epsilon^{abc}(q_1^{{\rm T}a}C\tilde{\Gamma}\tau q^b_2)h_v^c~,
j=\epsilon^{abc}(q_1^{{\rm T}a}C \Gamma_1\tau q^b_2)\Gamma_2 q_v^c~,
\end{array}
\end{equation}
where $\tilde{j}^v$ is the HQET baryonic current for the heavy quark $h_v$,
while $j$ is the protonic current. There are two choices both for
$\tilde{j}^v$ and $j$. We choose $\tilde{\Gamma}=\gamma_5$ for $\tilde{j}^v$
for the sake of simplicity since the numerical differences resulting from
the different choices of $\tilde {j}^{v}$ are not significant$\cite{liul}$.
We choose the tensor variant for $j$ since it seems to be more suitable for
studying the electromagnetic and strong properties of light baryons$\cite{eil}$, i.e.
\begin{equation}
\begin{array}{lll}
j&=&\epsilon^{abc} u^a C \sigma_{\mu\nu} u^b \sigma^{\mu\nu} \gamma_5 d^c \cr
&=&4 \epsilon^{abc}(u^a C \gamma_5 d^b u^c + u^a C d^b \gamma_5 u^c).
\end{array}
\end{equation} 

The three point correlator in present case is
\begin{equation}
\Pi (P^{'}, P, z)=i^2\int d^4xd^4ye^{ik\cdot x-iP\cdot y}
<0|T\tilde{j}^{v}(x)\bar{h}_{v}(0)\Gamma u(0)\bar{j}(y)|0>~,
\end{equation}
where $P^{'}=m_b v + k$ and $z= P\cdot v$.
After inserting a complete set of physical intermediate states,
as the phenomenological consequence of (4), we have 
\begin{equation}
\Pi(P^{'}, P, z)= f_{\Lambda_b} f_p {2 \over {(\omega - 2 \bar \Lambda)}}
        P_{+} \Gamma [F_1(z)+F_2(z) \not\! v ] {{\not\! P + m_p}\over{P^2-m_p^2}} + {\rm res},
\end{equation}
where $P_{+}=(1+\not\! v)/2$, $\bar{\Lambda}=m_{\Lambda_b}-m_b$, $\omega=2k \cdot v$ and $f_{\Lambda_b}$, $f_p$ are the so-called
"decay constants" which are given by
\begin{equation}
<0|\tilde{j}^v|\Lambda_b >=f_{\Lambda_b}u~, \\
<p(P)|\bar j|0>=f_p \bar u(P).
\end{equation}
They can be found in Ref.$\cite{liul}$ and Refs.$\cite{nrl,sn}$, respectively. To obtain
(5) we have taken into account (1) and the heavy quark limit.
  
Introducing the assumption of quark-hadron duality, the contribution
from higher resonant and continuum states can be treated as 
\begin{equation}
res.=\int_{D^{'}} d\nu ds{{\rho_{pert}(\nu,s,z)} \over{(\nu-\omega)(s-P^2)}}.
\end{equation}
The region $D^{'}$ is characterized by one or two continuum thresholds 
$\nu_c$, $s_c$. As a essential practice in QCD sum rule, in order to incorporate
the assumption, we should express the perturbative term in the form of 
dispersion relation
\begin{equation}
\Pi_{pert}^i (\omega,P^2,z)=\int d\nu ds{{\rho^i (\nu,s,z)}\over{(\nu-\omega)(s-P^2)}}, ~~~ (i=1,2),
\end{equation}
where $i=1,2$ denote the different
terms associated with $F_1$ and $F_2$, respectively.
 
The calculation of (4) can be done by using operator product expansion(OPE)
\begin{equation}
T[\tilde{j}^{v}(x)\bar{h}_{v}(0)\Gamma u(0)\bar{j}(y)]=\Sigma_n C_n (x,y) O_n (0).
\end{equation}
The first term represents the perturbative contribution, while the remaining
terms represent the nonperturbative effects after introducing the QCD vacuum
and condensates. Up to dimension 6, the relevant Feynman diagrams to compute (4) are plotted in
Fig.1. 

%%%%%%%%%%%%%%%%%%%%%%%figure%%%%%%%%%%%%%%%%%%%%%%%%%%%%
%\centerline{
%\includegraphics[height=11cm]{fig.ps}
%}
%{\footnotesize Fig.1  Feynman diagrams for the computation of $\Pi(P^{'}, P, z)$.
%The diagrams where the gluon condensates are generated from
%$b$-quark line are not plotted because the contributions corresponding to the diagrams
%are zero due to the use of fix-point gauge.}
%%%%%%%%%%%%%%%%%%%%%%%%figure%%%%%%%%%%%%%%%%%%%%%%%%

In our calculation, the coordinate representation is adopted. The heavy 
quark propagator
$$<T h_v(x)\bar h_v (0)>=\int_0^\infty dt \delta (x-t v) P_{+} $$
and the fixed point gauge$\cite{nsvz}$ are used. For our purposes it is sufficient
to retain the 
condensates with dimensions lower than 7. We use the following values of 
the condensates$\cite{isnr}$:
\begin{equation}
\begin{array}{rcl}
\langle\bar{q}q\rangle &\simeq&-(0.23~ {\rm GeV})^3~,\\
\langle\alpha_sGG\rangle &\simeq&0.04~ {\rm GeV}^4~.\\
%\langle g\bar{q}\sigma_{\mu\nu}G^{\mu\nu}q\rangle 
%&\equiv&m_0^2\langle\bar{q}q\rangle~, ~~~~~~m_0^2\simeq0.8~{\rm GeV}^2~.\\
\end{array}
\end{equation}
The normalization ${\rm Tr} \tau^{\dagger}\tau=1$ has been used in the
analysis.
In the fixed-point gauge, the space-time
translational invariance is violated, but it is restored by adding all
the diagrams in Fig. 1.  This is a check of our calculation.

In the standard way, we employ a double Borel Transformation $\omega\rightarrow
M, P^2\rightarrow T$ in order to suppress the higher excited states and 
continuum states contributions. The analytic expressions we got for $F_1$ and $F_2$
after Borel Transformation are:
\begin{equation}
\begin{array}{lll}
- 2f_{\Lambda_b}f_p F_1 e^{-2\bar{\Lambda}/M-m^2_p/T} &=&\int_0^{\nu_c}d\nu \int_{m_p^2}^{2\nu
z}ds \rho_{pert}^1 e^{-s/T-\nu/M}
- {1\over 3}\langle \bar q q \rangle^2 - \cr
& &{1\over {32\pi^4}}\langle\alpha_sGG\rangle \int_0^{T/4} (1-{{4\beta}\over T})  e^{-4\beta(
1-4\beta/T)/M^2-8\beta z/(TM)}d\beta,\cr
- 2f_{\Lambda_b}f_p m_p F_2 e^{-2\bar{\Lambda}/M-m^2_p/T} &=&\int_0^{\nu_c}d\nu \int_{m_p^2}^{2
\nu z}ds \rho_{pert}^2 e^{-s/T-\nu/M} + \cr
& & {1\over {8\pi^4}}\langle\alpha_sGG\rangle \int_0^{T/4}(1-{{4\beta}\over T}){\beta \over M
}e^{-4\beta(1-4\beta/T)/M^2-8\beta z/(TM)}d\beta,
\end{array}
\end{equation}
where
\begin{equation}
\begin{array}{lll}
\rho_{pert}^1&=&{1\over{32\pi^4\sigma^3}}[-2z^3\sigma^3-(-s+z(\nu+2z))^3+
3z^2(-s+z(\nu+2z))\sigma^2], \cr
\rho_{pert}^2&=&{-1\over{64\pi^4\sigma^3}}[s-2z^2+z(-\nu+\sigma)]^2[\nu s+8z^3-
4z^2(-2\nu+\sigma)-2z(-\nu^2+5s+\nu\sigma)],
\end{array}
\end{equation}
with $\sigma=\sqrt{-4s+(\nu+2z)^2}$. One can see from (11) that
the four-quark condensate represents the main nonperturbative
contribution to $F_1$, which is similar to the case of heavy to heavy transitions$\cite{huang}$.

In our numerical analysis, the "decay constants" and some other constants we used 
are$\cite{liul,nrl,data}$:
\begin{equation}
\begin{array}{rcl}
& &m_{\Lambda_b}=5.64 {\rm GeV}, m_p=0.938 {\rm GeV}, f_{\Lambda_b}=\sqrt{0.0003} {\rm GeV}^3\cr
& &f_p=0.0255 {\rm GeV}^3, \bar \Lambda=0.79 {\rm GeV}.
\end{array}
\end{equation}
It should be mentioned here that the values of $f_p$ defined in (6) is probably
different from
which we cited from Ref.$\cite{nrl}$ because we has adopted a different proton
current, that is, the "tensor current", instead of the "vector current"$\cite{nrl,sn}$.
However, if the approximations in sum rule calculations are justified to be good
 enough, these two currents should give roughly the same value of $f_p$. 
We find that with the threshold $\nu_c=3.7$GeV,
we can have a reasonably good window for $F_1$, where $1.4{\rm GeV}<{{4T}\over{m_b}}
=M<1.8{\rm GeV}$. The results are given in 
Fig.2 and Fig.3 respectively, where the different curves correspond to different
choices of the Borel parameters.

%%%%%%%%%%%%%%%%%%%figure%%%%%%%%%%%%%%%%%%%%%%%%%%%
%\begin{center}
%\includegraphics[height=12cm]{f1.eps}
%
%{\footnotesize Fig.2 Isgur-Wise function $F_1(z)$ with different values of Borel parameter
%T and M.}
%
%\includegraphics[height=12cm]{f2.eps}
%
%{\footnotesize Fig.3 Isgur-Wise function $F_2(z)$ with different values of Borel parameter
%T and M.}
%\end{center}
%%%%%%%%%%%%%%%%%%%%figure%%%%%%%%%%%%%%%%%%%%%%%%%%

The semileptonic decay $\Lambda_b\rightarrow p l \bar\nu$ can be analyzed directly after
we obtain the Isgur-Wise function $F_1$ and $F_2$. By neglecting the lepton mass (for
l=e,$\mu$), it is easy to show that the differential decay width is:
\begin{equation}
\begin{array}{lll}
{{d\Gamma}\over{dz}}&=&{{|V_{ub}|^2 G_F^2}\over{12\pi^3}}
               [ F_1^2 (3 m_p^2 z - 2 m_p^2 m_{\Lambda_b} + 3 m_{\Lambda_b}^2 z -4 m_{\Lambda_b} z^2) +
               2 F_1 F_2 (m_p^3 + 3 m_p m_{\Lambda_b}^2 - 6 m_{\Lambda_b} m_p z \cr
& &       + 2 m_p z^2) + F_2^2 (2 m_p^2 m_{\Lambda_b} - m_p^2 z + 
               3 m_{\Lambda_b}^2 z -8 m_{\Lambda_b} z^2+4 z^3)] 
               \sqrt{z^2-m_p^2}.
\end{array}
\end{equation}
The numerical results of ${{d\Gamma}\over{dz}}$ are shown in Fig.4. The total width
are $\Gamma=1.35\times 10^{-11}|V_{ub}|^2$GeV. The comparing values 
are $\Gamma=1.50\times 10^{-11}|V_{ub}|^2$GeV, when $M=1.4$GeV; 
$\Gamma=1.40\times 10^{-11}|V_{ub}|^2$, when$M=1.8$GeV. 

%%%%%%%%%%%%%%%%%%%%%%%%%figure%%%%%%%%%%%%%%%%%%%%%%%%%%
%\begin{center}
%\includegraphics[height=12cm]{c.eps}
%
%{\footnotesize Fig.4 The differential decay width with different values of Borel parameter.}
%\end{center}
%%%%%%%%%%%%%%%%%%%%%%%%%%figure%%%%%%%%%%%%%%%%%%%%%%%%%

There are a few comments to be made regarding the results and our computation:

(a) The choice of ${{4T}\over {m_b}}=M$ is purely for convenience. We find that
the max($T$, $M$) matters in the computation when the magnitudes of $T$ and $M$
are diverse greatly. The choice is somewhat similar to that
of Ref.$\cite{br}$, where the analyze of semileptonic decay of $B$ meson was performed.
The values of Borel parameters we used seem very large compared with that of 
Ref.$\cite{huang}$. In fact, we encountered the same situation mentioned in 
Ref.$\cite{dfnr}$, i. e., the influence of the four-quark condensate
is still large for the values of Borel parameters. Nevertheless, comparing to those in
Ref.$\cite{br}$, the values are similar. It seems to be natural that the Borel 
parameters in heavy to light transitions may take values different from those in
heavy to heavy transitions.
 
(b) Similar to the case in Ref.$\cite{nal}$, we find that $F_1$ is dominated by the 
four-quark condensate (about $63\%$) instead of the perturbative term. Therefore,  
to assume the hybrid sum rules$\cite{nal}$ may be better in order to calculate $F_1$.
 As for $F_2$, because it lacks the 
four-quark condensate, the perturbative term dominates . $F_2$ has no good stability
in the window. This is because the $\alpha_s$ corrections which are 
expected to be
more important for $F_2$ than for $F_1$ have not been included in this paper.
However, we could obtain a good
window for ${{d\Gamma}\over{dz}}$ and for the total width.

(c) The absolute values of $F_1$ and $F_2$ depend on
the "decay" constants of $\Lambda_b$ and $p$, which themselves have some 
uncertainties. The result for decay width will double these uncertainties.
Furthermore, there exists the uncertain CKM element $|V_{ub}|$ in the theoretical 
determination of the magnitude of decay width. In order to eliminate these 
uncertainties one considers the ratio $R=F_2/F_1$. Our numerical result for $R\simeq -0.42$ 
at zero recoil($z=m_p$) is in consistence with the experimental data for 
$\Lambda_b\rightarrow \Lambda e\bar\nu$$\cite{cleo}$. This is expected in both the heavy
quark limit and the light flavor $SU_3$ limit.
Moreover, our result suggests a tendency of the growth of $R$ with the 
final state baryon getting lighter. Note that
if the final state baryon is a heavy baryon, $R$ will approach zero. So the tendency
is in the right direction.

In summary, we have calculated the form factors of $\Lambda_b\rightarrow p$ in the
$m_b\rightarrow\infty$ limit from QCD sum rules within the framework of HQET. We
have also calculated the decay width of $\Lambda_b\rightarrow p l\bar{\nu}$ using
the obtained form factors.
With minor modifications, the results can be generalized to $\Lambda_c\rightarrow
\Lambda l \bar{\nu}$ which we shall analyze in detail elsewhere.

\vspace{0.5CM}

   We thank Y. B. Dai for discussions. C.F.Qiao wishes to thank the ICTP high energy 
physics group for the hospitality during his visit. This work is supported 
in part by the National Natural Science Foundation of China and the Hua Run Postdoctoral Science
Foundation of China.  

\vspace{2.0cm}

{\large \bf Figure Captions}
\vspace{1cm}

{Fig. 1 Feynman diagrams for the computation of $\Pi(P^{'}, P, z)$.
The diagrams where the gluon condensates are generated from
$b$-quark line are not plotted because the contributions corresponding to the diagrams
are zero due to the use of fix-point gauge.}
\vspace{1cm}

{Fig. 2 Isgur-Wise function $F_1(z)$ with different values of Borel parameter
T and M.}
\vspace{1cm}

{Fig. 3 Isgur-Wise function $F_2(z)$ with different values of Borel parameter
T and M.}
\vspace{1cm}

{Fig. 4 The differential decay width with different values of Borel parameter.}

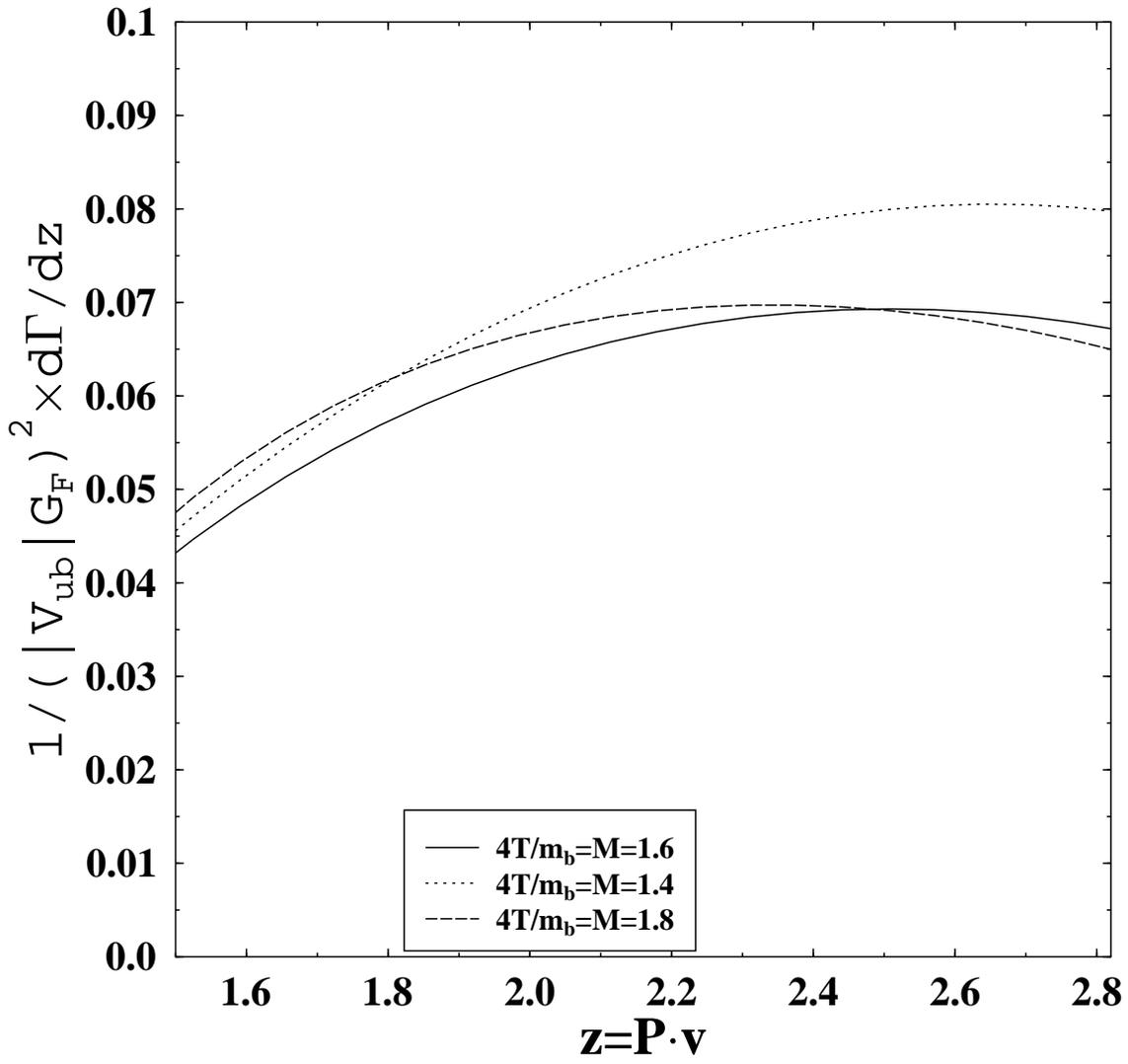
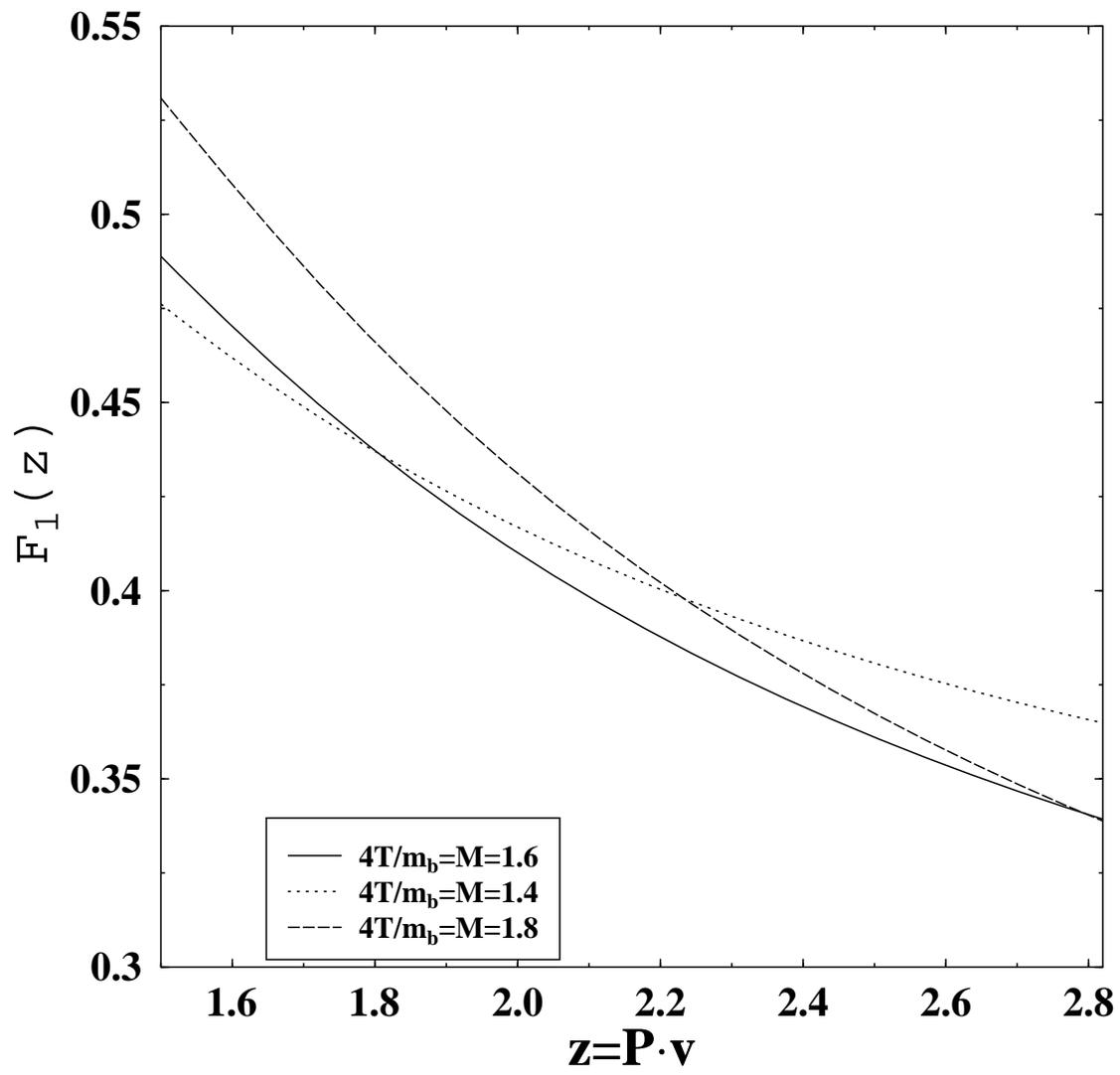
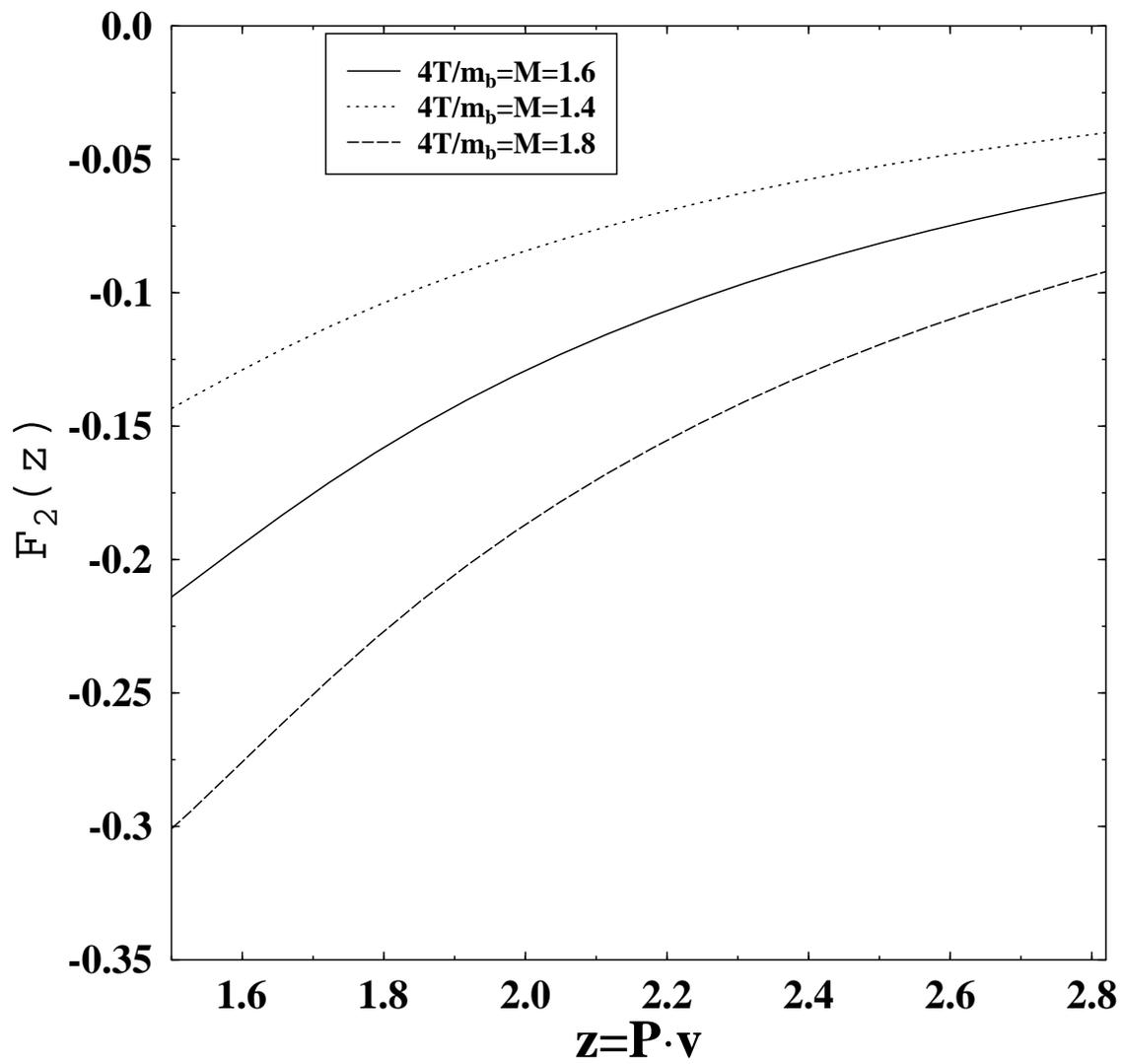
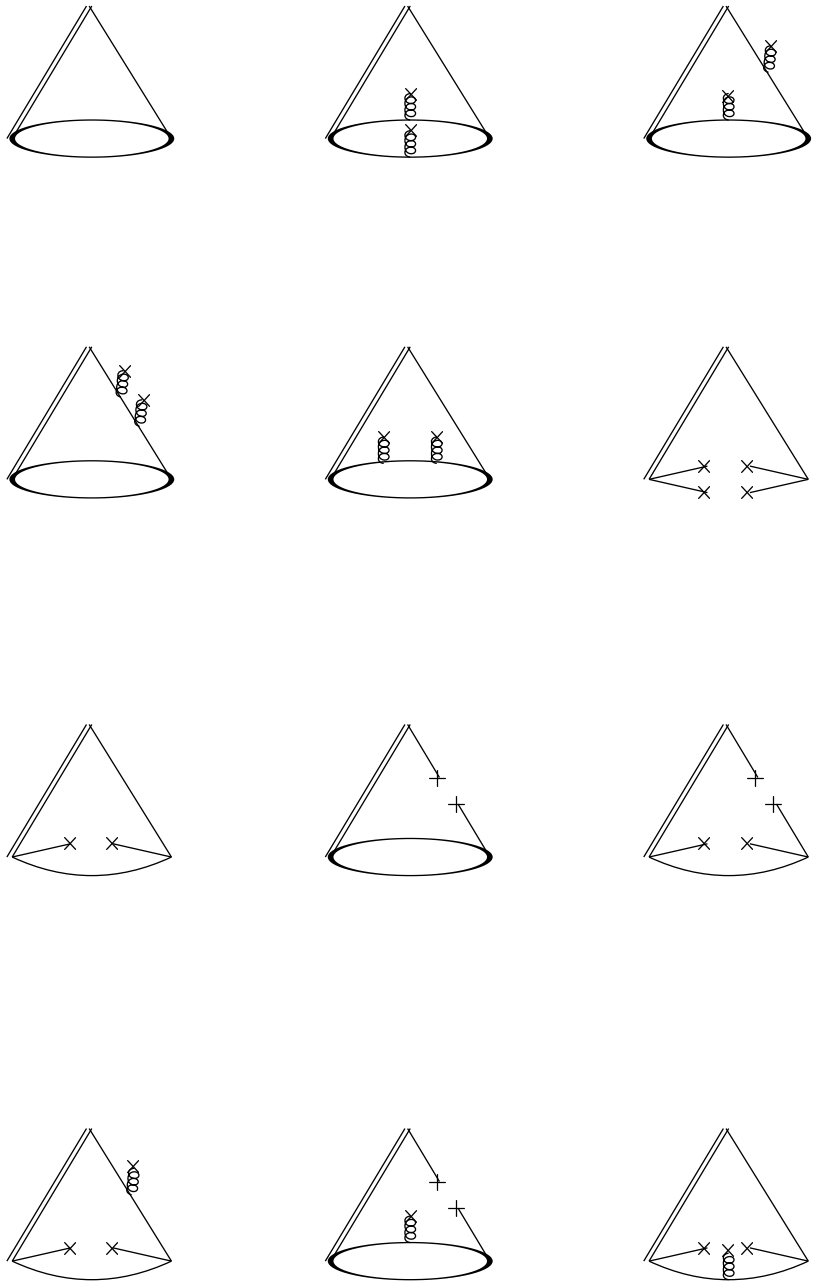
\end{document}